\documentclass[a4paper,11pt]{article}
\usepackage{pos}
\usepackage{subfigure}

\newcommand{\bs}{\boldsymbol}
\renewcommand{\vec}{\bs}

\usepackage{combelow}

\usepackage[normalem]{ulem}
\definecolor{limegreen}{RGB}{154,205,0}
\definecolor{brickred}{rgb}{0.8, 0.25, 0.33}
\definecolor{amethyst}{rgb}{0.6, 0.4, 0.8}
\definecolor{azure}{rgb}{0.0, 0.5, 1.0}
\definecolor{awesome}{rgb}{1.0, 0.13, 0.32}
\definecolor{purple}{rgb}{0.8,0,0.6}

\makeatletter
\DeclareFontFamily{OMX}{MnSymbolE}{}
\DeclareSymbolFont{MnLargeSymbols}{OMX}{MnSymbolE}{m}{n}
\SetSymbolFont{MnLargeSymbols}{bold}{OMX}{MnSymbolE}{b}{n}
\DeclareFontShape{OMX}{MnSymbolE}{m}{n}{
    <-6>  MnSymbolE5
   <6-7>  MnSymbolE6
   <7-8>  MnSymbolE7
   <8-9>  MnSymbolE8
   <9-10> MnSymbolE9
  <10-12> MnSymbolE10
  <12->   MnSymbolE12
}{}
\DeclareFontShape{OMX}{MnSymbolE}{b}{n}{
    <-6>  MnSymbolE-Bold5
   <6-7>  MnSymbolE-Bold6
   <7-8>  MnSymbolE-Bold7
   <8-9>  MnSymbolE-Bold8
   <9-10> MnSymbolE-Bold9
  <10-12> MnSymbolE-Bold10
  <12->   MnSymbolE-Bold12
}{}

\let\llangle\@undefined
\let\rrangle\@undefined
\DeclareMathDelimiter{\llangle}{\mathopen}%
                     {MnLargeSymbols}{'164}{MnLargeSymbols}{'164}
\DeclareMathDelimiter{\rrangle}{\mathclose}%
                     {MnLargeSymbols}{'171}{MnLargeSymbols}{'171}
\makeatother

\title{Moment of inertia and supervortical temperature of gluon plasma}

\author[a]{Victor~V.~Braguta}
\author[b,c]{Maxim~N.~Chernodub}
\author[d]{Ilya~E.~Kudrov}
\author*[a]{Artem~A.~Roenko}
\author[a,e]{Dmitrii~A.~Sychev}

\affiliation[a]{Bogoliubov Laboratory of Theoretical Physics, Joint Institute for Nuclear Research,  141980 Dubna, Russia}
\affiliation[b]{Institut Denis Poisson UMR 7013, Universit\'e de Tours,  37200 Tours, France}
\affiliation[c]{Nordita, Stockholm University, Roslagstullsbacken 23, SE-106 91 Stockholm, Sweden}
\affiliation[d]{Institute for High Energy Physics, NRC ``Kurchatov Institute'',  142281 Protvino, Russia}
\affiliation[e]{Moscow Institute of Physics and Technology,  141700 Dolgoprudny, Russia}

\emailAdd{vvbraguta@theor.jinr.ru}
\emailAdd{maxim.chernodub@univ-tours.fr}
\emailAdd{ilyakudrov@yandex.ru}
\emailAdd{roenko@theor.jinr.ru}
\emailAdd{sychev.da@phystech.edu}

\abstract{
Using lattice simulations, we analyze the influence of uniform rotation on the equation of state of gluodynamics. For a sufficiently slow rotation, the free energy of the system can be expanded into a series of powers of angular velocity. We calculate the moment of inertia given by the quadratic coefficient of this expansion 
using both analytic continuation and derivative methods, which demonstrate a good agreement between the results. 
We find that the moment of inertia unexpectedly takes a negative value below the ``supervortical temperature'' $T_s = 1.50(10) T_c$, vanishes at $T = T_s$, and becomes a positive quantity at higher temperatures. We discuss how our results are related to the scale anomaly and the magnetic gluon condensate. We point out that the negativity of the moment of inertia is in qualitative agreement with our previous lattice calculations, indicating that the rigid rotation increases the critical temperatures in gluodynamics and QCD.
}

\FullConference{The 40th International Symposium on Lattice Field Theory (Lattice 2023)\\
July 31st - August 4th, 2023\\
Fermi National Accelerator Laboratory\\}

\begin{document}
\maketitle

\section{Introduction}\label{sec:intro}
The droplets of strongly interacting quark-gluon plasma (QGP) with non-zero angular momentum are routinely produced in the non-central heavy-ion collisions~\cite{STAR:2017ckg, Becattini:2020ngo, Huang:2020dtn}. The measurements of global polarization of created $\Lambda,\bar\Lambda$-hyperons allow us to estimate an average vorticity of rotating QGP as $\omega$~$\approx (9 \pm 1)\times 10^{21} \, \mathrm{s}^{-1} \sim 0.03\, \mathrm{fm}^{-1} c \sim 7 \, \mathrm{MeV}$~\cite{STAR:2017ckg}.
Rotation may sufficiently affect the properties of QGP, which is possible to observe experimentally~\cite{Becattini:2020ngo, Huang:2020dtn}. 

The effects of relativistic rotation in field-theoretical systems are actively investigated within various theoretical approaches~\cite{Yamamoto:2013zwa, Braguta:2020biu, Braguta:2021jgn, Braguta:2021ucr, Braguta:2022str, Ambrus:2014uqa, Chen:2015hfc, Ambrus:2015lfr, Jiang:2016wvv, Chernodub:2016kxh, Chernodub:2017ref, Wang:2018sur, Zhang:2020hha, Chen:2020ath, Chernodub:2020qah, Fujimoto:2021xix, Golubtsova:2021agl, Sadooghi:2021upd, Jiang:2021izj, Chernodub:2022veq, Zhao:2022uxc, Chen:2022smf, Golubtsova:2022ldm, Braga:2023qej, Sun:2023dwh, Yang:2023vsw, Ambrus:2023bid, Mameda:2023sst} and the base tool of these studies is an assumption of rigid rotation.
This assumption is not realized in QGP produced in heavy-ion collisions, but it drastically simplifies analytical treatment and gives an opportunity to make predictions about effects in rotating QCD. 

At finite temperature, QCD exhibits a confinement-deconfinement transition and chiral transition.
The results of lattice simulations show that the critical temperatures of both transitions in rotating QCD increase with angular velocity~\cite{Braguta:2022str, Yang:2023vsw}. At the same time, the rotation in the fermionic sector tends to decrease the critical temperatures, but the contribution from gluons overcomes~\cite{Braguta:2022str}. It suggests that the gluon contribution is crucial for understanding the properties of rotating QCD. 
Similar conclusions also follow from some phenomenological models~\cite{Jiang:2021izj, Mameda:2023sst}.

Due to these reasons, we continue the detailed analysis of the gluon sector in rotating QCD.
In this paper, we summarize the main lattice results of our recent study, devoted to the mechanical properties and the equation of state of gluodynamics, presented in Refs.~\cite{Braguta:2023kwl, Braguta:2023yjn}.

\section{Free energy of rotating system and moment of inertia}\label{sec:free_energy}
A mechanical response of a thermodynamic ensemble to rigid rotation with the angular velocity $\vec{\Omega} = \Omega \vec e$ is described in terms of conjugated variable, the total angular momentum $\vec{J} = J\vec{e}$, which includes both spin and orbital contributions.\footnote{The vectors $\vec J$ and $\vec \Omega$ are parallel if the system rotates around one of its principal axes.}
In the co-rotating reference frame, the angular velocity is a natural variable of the free energy $F$ of the system. So, the total angular momentum can be expressed as follows
\begin{equation}\label{eq:J}
    \vec{J} = -\left(\frac{\partial F}{\vec{\partial \Omega}}\right)_T\, ,
\end{equation}
and the relation between the total angular momentum $\vec{J}  = I(T,\Omega) \vec{\Omega}$ and the angular velocity $\vec{\Omega}$ is given by the moment of inertia, which is the scalar quantity
\begin{equation}
	I(T,\Omega) = \frac{J(T,\Omega)}{\Omega} = - \frac{1}{\Omega} {\left( \frac{\partial F}{\partial \Omega} \right)}_{T}\,.
\label{eq:moment_inertia}
\end{equation}

For a classical system, the moment of inertia is determined by the mass (energy) distribution $\rho(T,x_\perp,\Omega)$, where $x_\perp$ is the distance to the rotational axis, and by the characteristic radius $R$ of the system.
Neglecting the effects of mass redistribution in the slowly rotating uniform system, $\rho(T,x_\perp,\Omega) \approx \rho_0(T)$, we can obtain well-known non-relativistic formula \begin{equation}\label{eq:I_classic}
    {I(T, \Omega)} = \int_V d^3 x \, x_\perp^2 \rho(T,x_\perp,\Omega) \simeq  \alpha \, {\rho_0(T)} V R^2 \equiv I(T,\Omega=0)\,,
\end{equation}
where the factor $\alpha$ depends on the geometry of the system.

The free energy of a rotating system may be represented as a series in angular velocity
\begin{equation}\label{eq:F_series}
F(T, V, \Omega) = F_0 (T,V) - \frac{F_2(T,V)}{2} \Omega^2 + \mathcal{O}(\Omega^4)\,,
\end{equation}
where $F_0(T,V) = -pV < 0$ is the free energy of non-rotating system.
As it follows from Eq.~\eqref{eq:moment_inertia}, the second coefficient $F_2$ in this expansion may be associated with the moment of inertia of rotating gluon plasma, $F_2 (T,V) = I (T,V,\Omega=0)$.
The moment of inertia is an extensive quantity, and using size-volume dependence of classical expression~\eqref{eq:I_classic}, we can rewrite Eq.~\eqref{eq:F_series} if the following form
\begin{equation}\label{eq:F_series_v}
    F(T, V, \Omega) = F_0(T,V) \left( 1 + \frac{K_2}{2} v_R^2 + \mathcal{O}(v_R^4) \right)\,,
\end{equation}
where $v_R = \Omega R$ is the linear velocity at some point at the boundary of the studied volume and the dimensionless coefficient $K_2 \equiv - I/(F_0 R^2)$ has a sense of a specific moment of inertia normalized by the free energy in the static case and transversal system size (note that $F_0 < 0$).
From the structure of Eq.~\eqref{eq:F_series_v}, we can conclude that the free energy of the rotating system should depend on the product $\Omega R$, but not on the radius of the system $R$ or angular velocity $\Omega$ itself.
Our lattice study of the critical temperature behavior in rotating gluodynamics shows that the prevail dependence of critical temperature is indeed on the boundary velocity squared~\cite{Braguta:2020biu, Braguta:2021jgn, Braguta:2021ucr, Braguta:2022str}, what supports this expectation. 

During the work with the rotating system, we should respect causality, which can be ensured by the appropriate maximal permissible value of boundary velocity $v_R$ in the used numerical setup.\footnote{This maximal value depends on the geometry of the system and choice of the boundary point,  which rotates with the linear velocity $v_R$. 
In our case of the cylinder with square cross-section $L_s\times L_s$ and $R = L_s/2$, we have a restriction $v_R < 1/\sqrt{2}$.}
For the rotating system, we also have an ambiguity with boundary conditions in the directions orthogonal to the rotational axis.
Although the boundary is screened due to the finite correlation length -- as confirmed by lattice simulations~\cite{Braguta:2021jgn} -- a few types of boundary conditions may be used,  
and this choice leads to a minor dependence of the results on boundary conditions.
Both the moment of inertia $F_2$ and the free energy of non-rotating system $F_0$ 
are affected by the boundaries, whereas in the dimensionless moment of inertia $K_2$, these dependencies are almost completely compensated. In this paper, we focused on the results with periodic boundary conditions.

\section{Lattice formulation of rotating QCD}\label{sec:lattice_formulation}
To perform Monte Carlo simulations of rotating gluodynamics, we follow the same approach as in Refs.~\cite{Yamamoto:2013zwa, Braguta:2020biu, Braguta:2021jgn, Braguta:2021ucr, Braguta:2022str}.
The action $S$ of the studied system should be formulated in the curved background of the co-rotating frame:
\begin{equation} \label{eq:g_E}
g^E_{\mu \nu} = 
\begin{pmatrix}
1 & 0 & 0 & x_2 \Omega_I \\
0 & 1 & 0 & -x_1 \Omega_I  \\ 
0 & 0 & 1 & 0 \\
x_2\Omega_I  & - x_1 \Omega_I & 0 & 1 + x_\perp^2 \Omega_I^2
\end{pmatrix}\,,
\end{equation}
written in the Euclidean coordinates $x^\mu = (x^1,\dots,x^4)$, where $x_4 = - i t$ is the imaginary time coordinate, and $x^2_\perp = x^2_1 + x_2^2$.
The system rotates around the $x_3$ axis.  
The angular velocity in Eq.~\eqref{eq:g_E} is put in the purely imaginary form $\Omega_I = - i \Omega$ to avoid the sign problem~\cite{Yamamoto:2013zwa}. In particular, the velocity $v_R=\Omega R$ at the boundary point with the distance $R$ to the rotational axis becomes imaginary $v_I = - i v_R$.
In the result, the Euclidean gluon action has the following form:
\begin{align}
    S = \frac{1}{4 g_0^2} \int\! d^4 x\, \sqrt{g_E}\,  g_E^{\mu \nu} g_E^{\alpha \beta} F_{\mu \alpha}^{a} F_{\nu \beta}^{a}
    \equiv S_0 + S_1 \Omega_I + S_2 \frac{\Omega_I^2}{2}\, ,
    \label{eq:Sg_general}
\end{align}
where $S_0$ is the action of gluons without rotation, and $S_1$, $S_2$ are some operators, the explicit form of which can be found in Ref.~\cite{Braguta:2023kwl}. In our lattice simulations, the last two terms of the action~\eqref{eq:Sg_general}, which are coupled with the rotation, are discretized following Refs.~\cite{Yamamoto:2013zwa, Braguta:2021jgn}, whereas for the non-rotating part $S_0$ we use the tree-level improved Symanzik gauge action~\cite{Curci:1983an}.

\section{Results of lattice simulations with imaginary angular velocity}\label{sec:lattice_imag_results}
We calculate the free energy density $f = F/V$ of the rotating system 
using the standard relations~\cite{Boyd:1996bx} 
\begin{subequations}\label{eq:F_S_lat}
\begin{align}
    \frac{f(T)}{T^4} & = - N_t^4 \int_{\beta_0}^{\beta} d \beta' \Delta s(\beta')\,,
    \label{eq:F_lattice} \\
    \Delta s(\beta) & = \langle s(\beta) \rangle_{T=0} - \langle s(\beta) \rangle_{T},
    \label{eq:Delta_S}
\end{align} 
\end{subequations}
where the $\mathrm{SU}(N_c)$ lattice coupling $\beta = 2 N_c/g_0^2$ is expressed in terms of the bare continuum coupling constant $g_0$, while $s=T/V\cdot S$ is the density of the lattice action $S$.
The calculations are performed on the lattices of size $N_t \times N_z \times N_s^2 = N_t \times 40 \times 41^2 $ with $N_t = 5, 6, 7, 8$ and $N_t = 40$ for the zero temperature subtraction in Eq.~\eqref{eq:Delta_S}. The imaginary velocity at the boundary is identified with the velocity at the middle of the boundary side, $v_I = \Omega_I R$, where ${R = a(N_s - 1)/2}$ is the distance from the boundary to the rotational axis. Along the integration path in Eq.~\eqref{eq:F_lattice}, the velocity $v_I$ remains fixed, corresponding to the lines of constant physics~\eqref{eq:F_series_v}.
In this study, the linear velocity takes the following values: $v_I^2 = 0.000, 0.015, 0.030, 0.045, 0.060, 0.075, 0.090$, i.e., $v_I^2 \ll 1$.

\begin{figure}[!th]
\subfigure[]{\label{fig:fT4-I}
\includegraphics[width=.485\textwidth]{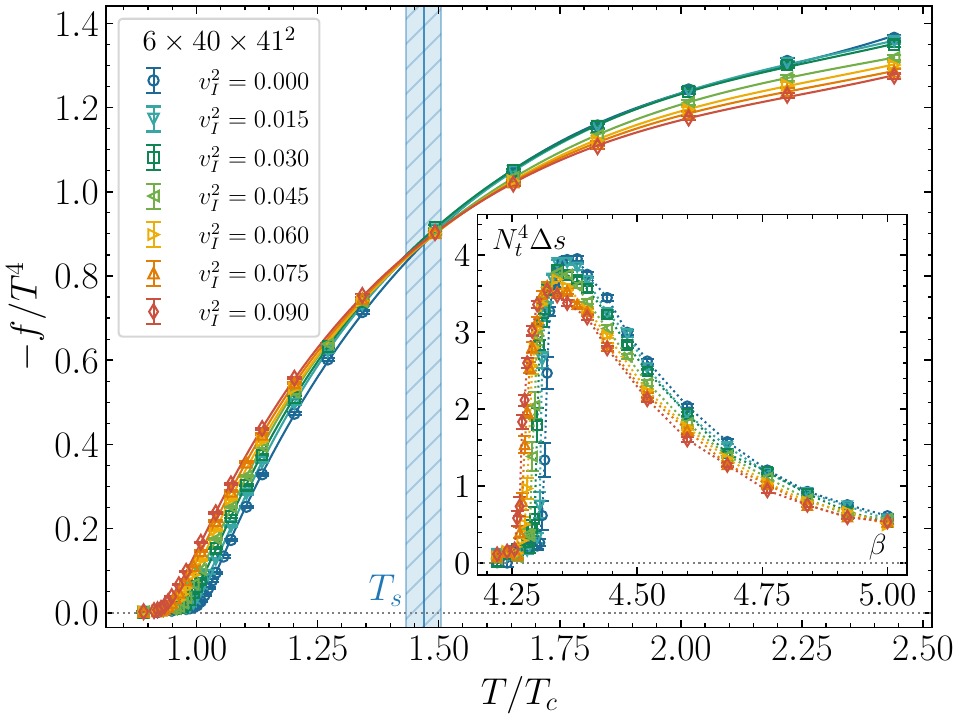}
}
\hfill
\subfigure[]{\label{fig:K2-I}
\includegraphics[width=.485\textwidth]{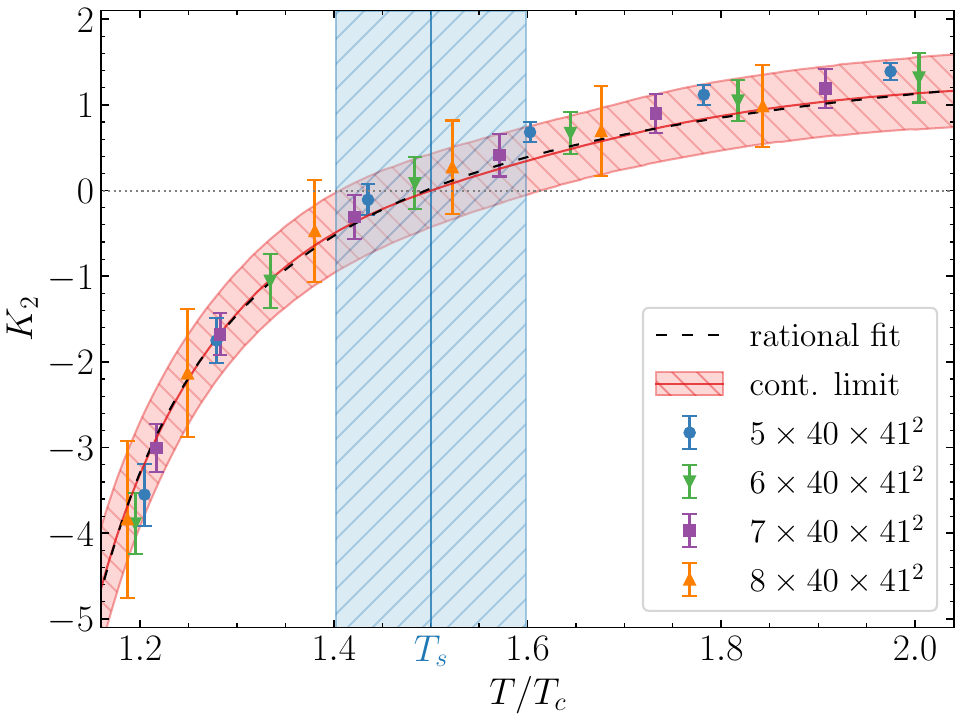}
}
\vspace{-0.5em}
\caption{
\subref{fig:fT4-I}: The free energy density $f$ in the co-rotating frame as a function of the temperature $T$ for the {$N_t = 6$} lattice.
The vertical line shows the supervortical temperature $T_s$ for this lattice.
The inset shows the expectation value of the lattice action density $\Delta s$ as a function of the lattice gauge coupling~$\beta$.
Both plots are given for several imaginary velocities squared $v_I^2$ at the boundary of the system.
\subref{fig:K2-I}: The dimensionless specific moment of inertia $K_2$ of the gluon plasma as a function of the temperature $T$. The red-shaded region, with the central values marked by the red solid line, denotes the continuum extrapolation. The position of the supervortical temperature $T_s$ in the continuum limit is marked by the vertical line, which separates the unstable ($T < T_s$) and stable ($T > T_s$) regimes of rigid plasma rotation. The figures are from Ref.~\cite{Braguta:2023yjn}.}\label{fig:2303}
\end{figure}

In the inset of Fig.~\ref{fig:fT4-I}, we show the normalized difference of lattice action densities~$\Delta s$, Eq.~\eqref{eq:Delta_S}, which enters the free energy density~\eqref{eq:F_lattice}. At zero velocity of the rotation, we recover the known result for gluodynamics~\cite{Boyd:1996bx}. As the imaginary velocity $v_I$ increases, the transition shifts towards smaller lattice couplings $\beta$, signaling that the critical temperature $T_c = T_c(v_I)$ decreases as the {\it imaginary} angular frequency $\Omega_I$ (the velocity $v_I$ of the rotation) raises. This result is in agreement with previous lattice calculations~\cite{Braguta:2020biu, Braguta:2021ucr, Braguta:2021jgn, Braguta:2022str}. 

The normalized free energy density in the co-rotating frame, $ - f/T^4$, calculated via Eq.~\eqref{eq:F_lattice}, is shown in Fig.~\ref{fig:fT4-I}.
This quantity is a monotonically raising function of the temperature $T$ at all imaginary velocities $v_I$, indicating the presence of a plateau at large $T$ for each fixed~$v_I$. 
One can see from Fig.~\ref{fig:fT4-I} that the free energy density $f$ is a rising (diminishing) function of $v_I^2$ at fixed temperature $T < T_s$ ($T > T_s$). We fit the free energy density  with a parabolic function of $v_I$:
\begin{align} \label{eq:f_fit_vI}
    f(T,v_I)  = f_0(T) \Bigl(1 - \frac{1}{2} K_2(T) v_I^2 \Bigr)\,,
\end{align}
where $f_0$ and $K_2$ serve as fit parameters (with $f_0 < 0$).
This ansatz corresponds to Eq.~\eqref{eq:F_series_v} after the analytic continuation procedure $v_I^2 = - v_R^2$. The results for a dimensionless specific moment of inertia $K_2$ are shown in Fig.~\ref{fig:K2-I} for the used lattices and in the continuum limit ($a\to 0$, or, equivalently, $1/N_t\to 0$ at a fixed temperature $T$). 

A striking feature of the free energy, Fig.~\ref{fig:fT4-I}, is that the curves corresponding to different $v_I$ intersect at the same ``supervortical'' temperature $T_s$.
It signalizes that at this temperature, the free energy~\eqref{eq:f_fit_vI} loses, at least for slow rotation $v_I^2 \ll 1$, the dependence on the rotational frequency.
Therefore, the rigidly rotating gluon plasma loses its moment of inertia at $T = T_s$.
We use the $K_2(T_s) = 0$ property as a definition of the supervortical temperature, and show its continuum limit value $T_s = 1.50(10)\, T_c$ in Fig.~\ref{fig:K2-I} using the vertical line.\footnote{The results for the system with open boundary conditions give a consistent value $T_s = 1.53(15)\,T_c$.}

The continuum limit of the dimensionless moment of inertia $K_2$ can be well reproduced by a rational function  $K_2^{\mathrm{(fit)}}(T) = K_2^{(\infty)} - {c}/{(T/T_c -1)}$, where the best-fit parameters include the high-temperature asymptotic value $K_2^{(\infty)} = 2.23(39)$. 
Note that the high-$T$ limit of the moment of inertia is a non-universal quantity that may depend on the geometry of the system. 

\section{Results of lattice simulations at zero angular velocity}\label{sec:lattice_static_results}
Using Eqs.~\eqref{eq:moment_inertia},~\eqref{eq:F_series},~\eqref{eq:Sg_general}, we can represent the moment of inertia in the following form
\begin{equation}\label{eq:I_zero_omega}
    I = F_2 = T \frac {\partial^2 {\log Z}} {\partial \Omega^2} \biggl |_{\Omega=0} 
    =  {T \left( \llangle S_1^2 \rrangle_T - \llangle S_1 \rrangle_T^2 + \llangle S_2 \rrangle_T \right)}\,,
\end{equation}
where $\llangle{\mathcal O}\rrangle_T = \langle {\mathcal O} \rangle_T - \langle {\mathcal O} \rangle_{T=0}$ denotes the thermal part of the expectation value of an operator~${\mathcal O}$. The moment of inertia may be independently computed on the lattice directly by Eq.~\eqref{eq:I_zero_omega}, and only the simulation at zero angular velocity is needed in this case.
The simulation may be conducted with convenient periodic boundary conditions in all directions, allowing us to make an additional averaging over the position of the rotational axis during the calculation of the operators in Eq.~\eqref{eq:I_zero_omega}.

\begin{figure}[!th]
\subfigure[]{\label{fig:456}
\includegraphics[width=.485\textwidth]{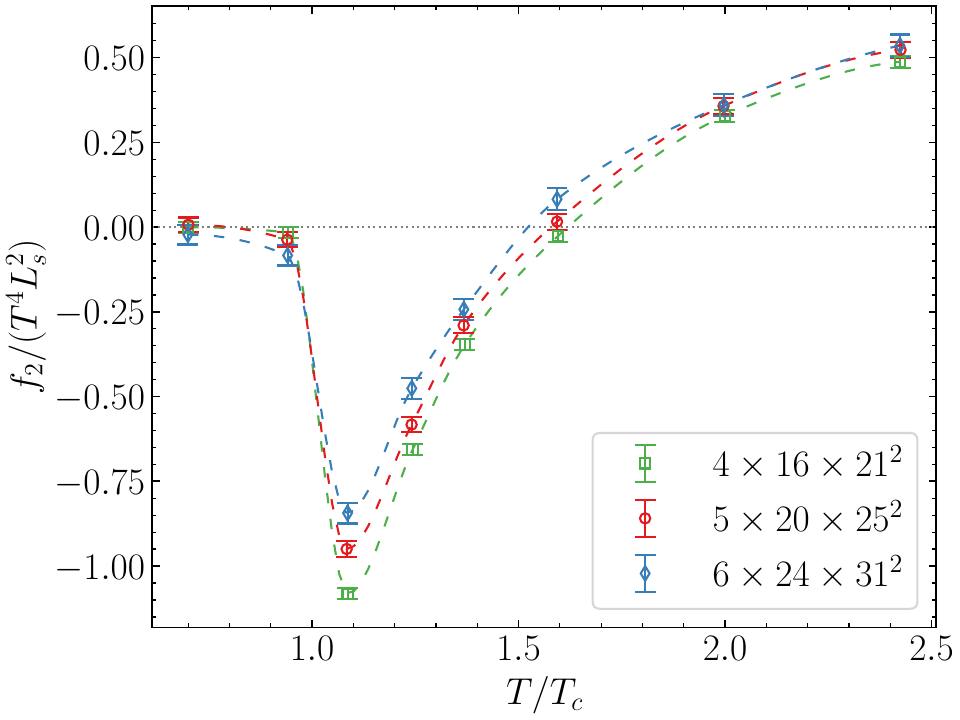}
}
\hfill
\subfigure[]{\label{fig:5zs}
\includegraphics[width=.485\textwidth]{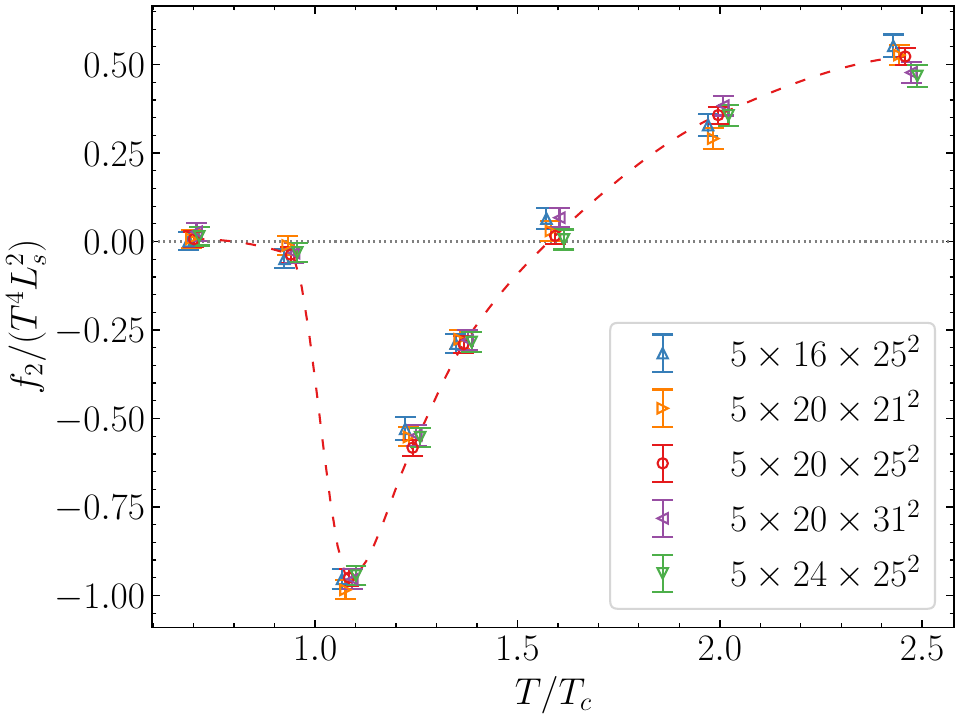}
}
\vspace{-0.5em}
\caption{
The ratio $f_2/(T^4 L_s^2)$ as a function of temperature (in units of the deconfinement critical temperature) for the lattices with different lattice spacing~\subref{fig:456} and different spatial extensions in $xy$- and $z$- directions~\subref{fig:5zs}.
The figures are from Ref.~\cite{Braguta:2023kwl}.}
\label{fig:JETPL}
\end{figure}

In order to keep under control the finite spacing effects and take the continuum extrapolation of the results, we perform the calculations on the lattices $4\times 16\times 21^2$, $5\times 20\times 25^2$, $6\times 24\times 31^2$.\footnote{Due to the structure of studied operators, statistical errors grow quite fast with the volume of the system, forcing us to abandon lattices with larger $N_t$ in this study.}
For the zero-temperature subtraction in Eq.~\eqref{eq:I_zero_omega}, we also compute the same operators on the lattices with large temporal extension $N_t = N_z$.
In the Fig.~\ref{fig:456} we present the temperature dependence of the ratio $f_2/(T^4 L_s^2)$, where we introduce the notations $f_2 = F_2/V$ and $L_s = a N_s$, for different~$N_t$. This ratio has a sense of the specific moment of inertia (compare Eqs.~\eqref{eq:I_classic},~\eqref{eq:I_zero_omega}), like $K_2$, but normalized by the temperature in fourth power $T^4$.

The specific moment of inertia is a non-extensive quantity, and it should be independent of the system sizes. To check this property, we make the simulations on the lattices with different extensions in $xy$- and $z$- directions (i.e., orthogonal to and along the rotational axis).
The results are presented in Fig.~\ref{fig:5zs}. From Fig.~\ref{fig:JETPL}, one can see that the moment of inertia of gluon plasma is close to zero in the confinement phase, it is negative in the temperature region $T_c \lesssim T < T_s$, where $T_s \sim 1.5 T_c$, and it is positive at the higher temperatures $T > T_s$, where the results should be consistent with the Stefan–Boltzmann limit and perturbative calculations.

\begin{figure}[!th]
\centering
\includegraphics[width=.485\textwidth]{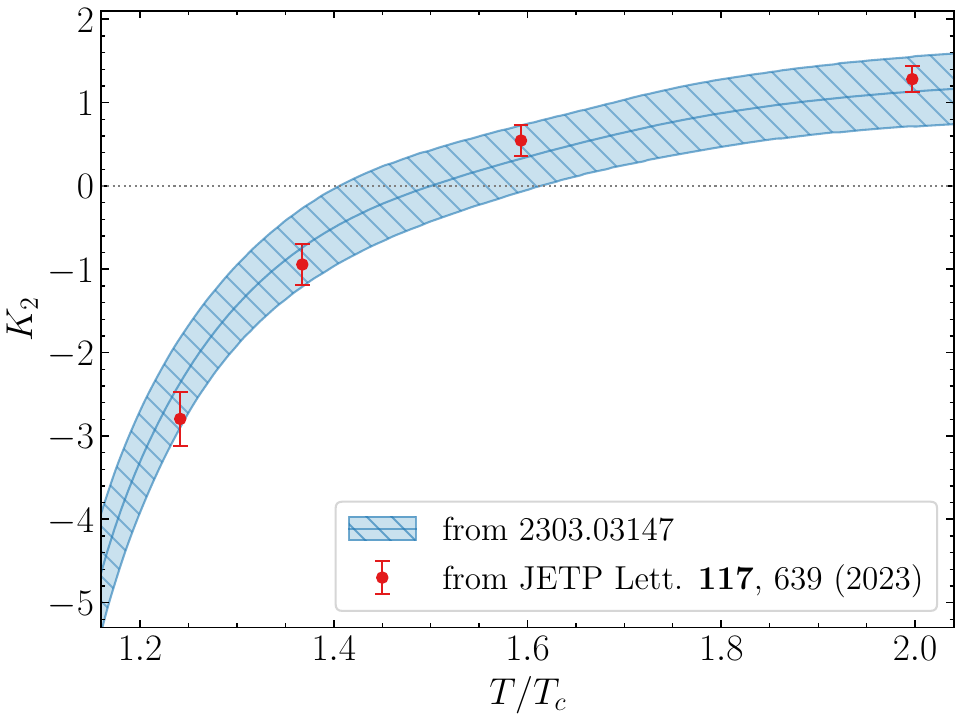}
\caption{
The  dimensionless  specific moment of inertia $K_2$ of the gluon plasma as a function of the temperature $T$,
obtained after the analytic continuation procedure (the blue-shaded region, from Ref.~\cite{Braguta:2023yjn}), and independent results of the direct computation on the non-rotating lattices (the red points, from Ref.~\cite{Braguta:2023kwl}).
}\label{fig:K2-comp}
\end{figure}

Using the data from Fig.~\ref{fig:JETPL}, we can take continuum extrapolation results for the dimensionless moment of inertia $K_2$ from static lattices and compare it with the results from Sec.~\ref{sec:lattice_imag_results}, obtained from the simulations with imaginary angular velocity. The continuum limit results of these two independent approaches (see Fig.~\ref{fig:K2-comp}) show a good agreement with each other. This coincidence supports the validity of  our analytic continuation procedure and reinforces the main claim about the negativity of the moment of inertia made in our papers~\cite{Braguta:2023kwl,Braguta:2023tqz,Braguta:2023yjn}.

\section{Moment of inertia and magnetic gluon condensate}
The expression for the moment of inertia~\eqref{eq:I_zero_omega} may be rewritten in a form that has a more transparent physical meaning.
Using the exact form of the operators $S_1$, $S_2$, we get~\cite{Braguta:2023yjn} 
\begin{equation}\label{eq:I_two_terms}
    I = I_{\textrm{mech}} + I_{\textrm{magn}}\,.
\end{equation}
The ``mechanical'' term $I_{\textrm{mech}}$ in Eq.~\eqref{eq:I_two_terms} describes the fluctuations of the total angular momentum of gluons at $\Omega = 0$  via a standard linear-response form
\begin{equation} \label{eq:I_mech}
    {I_{\text{mech}}} = \frac{1}{T} \left( \left\llangle (J^3)^2 \right\rrangle_T - \left\llangle J^3 \right\rrangle_T^2 \right)\,, \qquad\qquad
    J^i = \frac{1}{2} \int_V d^3 x \, \epsilon^{ijk} M^{jk}_{0}({\bs x})\,, 
\end{equation}
where $M^{ij}_{0}({\bs x}) = x^i T^{j0}({\bs x}) - x^j T^{i0}({\bs x})$, and $T^{\mu\nu}$ is the gluon stress-energy tensor. This contribution to the moment of inertia is strictly positive $I_{\text{mech}} > 0 $. The second term in Eq.~\eqref{eq:I_two_terms}, ``magnetic'' term, is closely related to the magnetic part of thermal gluon condensate~\cite{Braguta:2023yjn}:
\begin{equation} \label{eq:I_magn}
    I_{\text{magn}} = \frac{1}{3}\int_V d^3 x \, x_\perp^2 \left\llangle{(F_{ij}^{a})^2}\right\rrangle_{T} = \frac{\alpha}{3} V R^2 \left\llangle{(F_{ij}^a)^2}\right\rrangle_T\,.
\end{equation}
This formula has the same form as the classical expression for the moment of inertia~\eqref{eq:I_classic}, where the mass density is $\rho_0 = \llangle{(F_{ij}^a)^2}\rrangle_T/3$.
The total thermal gluon condensate $\llangle{(F_{\mu\nu}^a)^2}\rrangle_T$, which is negative at any temperature, is related to the scale anomaly: $\langle T^\mu_\mu \rangle = - \llangle{(F_{\mu\nu}^a)^2}\rrangle_T > 0$. But its magnetic part $\llangle{(F_{ij}^a)^2}\rrangle_T$ reverses the sign at $T\sim 2 T_c$, and becomes positive at higher temperatures~\cite{Boyd:1996bx}. Collecting both contributions in Eq.~\eqref{eq:I_two_terms} we obtain the negative values of the moment of inertia below ``supervortical'' temperature $T_s = 1.50(10) T_c < 2 T_c$, which agrees with the behavior of magnetic gluon condensate.

The division of the moment of inertia into two parts will also be valid in QCD.
Fermionic action in a rotating reference frame contains terms that are coupled with angular velocity, but they are only linear by $\Omega$~\cite{Yamamoto:2013zwa, Braguta:2022str}, so, they should be added to $S_1$ in Eq.~\eqref{eq:I_zero_omega}. These fermionic terms have a sense of orbital momentum and spin of quarks, and they make a positive contribution to the mechanical part $I_{\text{mech}}$, while the gluomagnetic part $I_{\text{magn}}$ stays negative in QCD. Thus, we believe that the moment of inertia of quark-gluon plasma is also negative in a region near $T_c$.

\section{Discussion and conclusions}\label{sec:conclusions}
Using numerical lattice simulations, we analyze the effects of rigid rotation on the equation of state of gluodynamics. We calculate the free energy of a rotating system for a set of fixed (imaginary) velocities of rotation, as well as the quadratic coefficient $F_2$ of the free energy expansion into a series of powers of angular velocity. This coefficient is associated with the isothermal moment of inertia of gluon plasma. We also calculate the moment of inertia using the direct derivative method at zero angular velocity and observe a good agreement with the results of the first method, which is based on the analytic continuation procedure.

The moment of inertia unexpectedly takes a negative value below the ``supervortical'' temperature $T_s = 1.50(10)\, T_c$, vanishes at $T = T_s$, and becomes a positive quantity at higher temperatures.
The negative moment of inertia originates from the thermal part of magnetic gluon condensate, which also takes negative values in the vicinity of the critical temperature. The results indicate a thermodynamic instability of gluon plasma with respect to rigid rotation below the ``supervortical'' temperature (see discussion in Ref.~\cite{Braguta:2023yjn}). The possible reason for this instability may lie in the peculiarities of the spin-orbital coupling for gluons in this temperature range, which causes the negative Barnett effect implying a negative spin-vorticity coupling~\cite{Braguta:2023tqz}.

We argue that these features of the gluon sector should also change the properties of quark-gluon plasma in the same way. Our results also suggest that the puzzling discrepancy between lattice~\cite{Braguta:2020biu, Braguta:2021ucr, Braguta:2021jgn, Braguta:2022str, Yang:2023vsw} and analytical~\cite{Jiang:2016wvv, Chernodub:2016kxh, Chernodub:2017ref, Wang:2018sur, Zhang:2020hha, Sadooghi:2021upd, Chen:2020ath, Zhao:2022uxc, Chernodub:2022veq, Fujimoto:2021xix, Golubtsova:2021agl, Chen:2022smf, Golubtsova:2022ldm, Chernodub:2020qah, Braga:2023qej, Sun:2023dwh} predictions for the critical temperature of rotating (quark-)gluon plasma might originate from the scale anomaly which should be taken into account appropriately, because the magnetic gluon condensate plays a key role in rotating QCD.

\acknowledgments
The authors are grateful to Oleg Teryaev and Victor Ambru\cb{s} for useful discussions and to Andrey Kotov for his participation at the early stage of the project.
This work has been carried out using computing resources of the Federal collective usage center Complex for Simulation and Data Processing for Mega-science Facilities at NRC ``Kurchatov Institute'', \url{http://ckp.nrcki.ru/}  and the Supercomputer  ``Govorun'' of Joint Institute for Nuclear Research. The work of VVB, IEK, AAR, and DAS, which consisted in the lattice calculation of the observables used in the paper,  was supported by the Russian Science Foundation (project no. 23-12-00072). MNC is thankful to the members of Nordita for their kind hospitality. 

\bibliographystyle{JHEP}
\bibliography{biblio.bib}

\end{document}